\documentclass[11pt,twoside]{article}
\usepackage{macro-fsut-eng}
\usepackage{graphicx}

\usepackage[T1]{fontenc} 

\usepackage{latexsym}
\usepackage{verbatim}

\newcommand\bfx{{\mathbf x}}
\newcommand\bfk{{\mathbf k}}
\newcommand{\dn}{\delta N}

\begin{document}

\vskip 1.0cm
\markboth{Y.~Rodríguez et al.}{Towards a proof of the equivalence between FRW expansion and statistical isotropy}
\pagestyle{myheadings}

\hspace{9cm}
PI/UAN-2014-581FT

\vspace*{0.5cm}
\title{Towards a proof of the equivalence between FRW background expansion and statistical isotropy}

\author{Yeinzon~Rodríguez,$^{1,2}$ L.~Gabriel~Gómez,$^1$ and Carlos~M.~Nieto$^2$}
\affil{$^1$Centro de Investigaciones en Ciencias Básicas y Aplicadas, Universidad Antonio Nariño, Cra 3 Este \# 47A - 15, Bogotá D.C. 110231, Colombia\\
$^2$Escuela de Física, Universidad Industrial de Santander, Ciudad Universitaria, Bucaramanga 680002, Colombia}

\begin{abstract}
We will expose in this paper our advances towards a proof of the equivalence between FRW background expansion, during some period of time that contains primordial inflation, and the statistical isotropy of the primordial curvature perturbation $\zeta$ at the end of this period of time.  Our motivation rests on the growing interest in the existence of a preferred direction in the Universe hinted by the continuous presence of anomalies in the CMB data. \end{abstract}

\section{Introduction}

Cosmology is based on a, once believed, sacred principle:  the homogeneity and isotropy at large scales.  The actual meaning of this statement is that perturbations in the energy density distribution at large scales, and in the CMB temperature, are statistically homogeneous and isotropic, i.e., their $n$-point correlators in real space are invariant under spatial translations and spatial rotations. 
However, the CMB data releases have consistently presented a possible indication of the existence of a preferred direction in the Universe in relation with the different anomalies in the data (Ade et. al., 2014; Bennett et. al., 2013).  Names as ``the axis of evil'' employed to describe some of the anomalies are evidence of both the idea of a preferred direction and the position of the cosmologists community about the violation of the sacred principle.  There are two ways of thinking of a preferred direction in the Universe, one at the background level and the other at the perturbative level:  at the background level we can have non-FRW expansion which can be associated, for instance, to the presence of a shear in the metric or an anisotropic curvature;  whereas at the perturbative level we can have statistical anisotropy in the perturbations, especially in the primordial curvature perturbation $\zeta$.  It is quite reasonable to think that a non-FRW background expansion always feeds the statistical anisotropy;  moreover, sometimes a FRW background expansion is assumed in the analysis of the statistical anisotropy.  Notwithstanding, there is no any proof yet regarding the actual relation between these two characteristics.  It is our mission in this paper to expose our advances towards such a kind of proof.  We will sketch the main argument of a proof of the equivalence between FRW background expansion, during some period of time that contains primordial inflation, and the statistical isotropy of the primordial curvature perturbation $\zeta$ at the end of this period of time, at the level of the power spectrum, leaving the most delicate points for a deeper analysis and discussion in a future publication.

\section{Statistical homogeneity and isotropy}

Quantum mechanics only allows us to predict probabilities of different outcomes after an experiment in an ensemble of systems, in contrast to classical mechanics which does allow us to predict the exact outcome after an experiment in just one element of the ensemble.  Since the underlying physical mechanism in the generation of cosmological perturbations is of quantum nature, the cosmologists are more interested in studying the statistical properties of a perturbation map, say the CMB map or the galaxy distribution map. The way of doing this is via the $n$-point correlators of the perturbations in real space.  Let's define a scalar cosmological perturbation $\beta({\bf x})$ in real space and make a Fourier integral expansion
\begin{equation}
\beta({\bf x}) \equiv \int \frac{d^{3} k}{(2 \pi)^{3}} e^{i {\bf k \cdot x}}\beta({\bf k}) \,, \label{ffe}
\end{equation}
where $\beta({\bf k})$ is the Fourier mode function of $\beta({\bf x})$. The $n$-point correlators of $\beta({\bf x})$ are averages over the ensemble of the products $\beta({\bfx_1}) \beta({\bfx_2}) ... \beta({\bfx_n})$ where ${\bfx_1}$, ${\bfx_2}$, ..., ${\bfx_n}$ represent different points in space\footnote{The ensemble average inside the integral is over the Fourier mode functions only since they are the stochastic variables.}:
\begin{eqnarray}
\langle \beta({\bfx_1}) \beta({\bfx_2}) ... \beta({\bfx_n}) \rangle \equiv \int \frac{d^{3} k_1}{(2 \pi)^{3}} \frac{d^{3} k_2}{(2 \pi)^{3}} ... \frac{d^{3} k_n}{(2 \pi)^{3}} && e^{i ({\bfk_1 \cdot \bfx_1} + {\bfk_2 \cdot \bfx_2} + ... + {\bfk_n \cdot \bfx_n})} \times \nonumber \\
&& \times \langle \beta({\bfk_1}) \beta({\bfk_2}) ... \beta({\bfk_n}) \rangle \,. \label{expan}
\end{eqnarray}
Thus, the correlation functions in real space may be studied via the correlation functions in momentum space.  Let's see now the meaning of statistical homogeneity and statistical isotropy.

\vspace{3pt}
\noindent {\bf Statistical homogeneity:}  Of course the perturbation map is not homogeneous (i.e., it is not invariant under spatial translations), but it may be that the probability distribution function governing $\beta({\bf x})$ is, which is called statistical homogeneity. This means that the $n$-point correlators in real space are invariant under translations in space, i.e.
\begin{equation}
\langle \beta({\bfx_1} + {\bf d}) \beta({\bfx_2} + {\bf d}) ... \beta({\bfx_n} + {\bf d}) \rangle = \langle \beta({\bfx_1}) \beta({\bfx_2}) ... \beta({\bfx_n}) \rangle \,,
\end{equation}
where ${\bf d}$ is some vector in real space establishing the amount of spatial translation. The only way of achieving this, in view of Eq. (\ref{expan}), is expressing the argument in the exponential function inside the integral as the addition of several terms of the form $f({\bf x}_i - {\bf x}_j)$, which in turn is possible (but it is not the only possibility) if the $n$-point correlators in momentum space are proportional to a Dirac delta function:
\begin{equation}
\langle \beta({\bfk_1}) \beta({\bfk_2}) ... \beta({\bfk_n}) \rangle \equiv (2\pi)^3 \delta^{3}({\bfk_{12...n}}) M_\beta ({\bfk_1}, {\bfk_2}, ... , {\bfk_n}) \,. \label{shcond}
\end{equation}
In the previous expression, ${\bfk_{12...n}}$ means ${\bfk_1} + {\bfk_2} + ... + {\bfk_n}$, and the function $M_\beta ({\bfk_1}, {\bfk_2}, ... , {\bfk_n})$ is called the $(n-1)$-spectrum.  

\vspace{3pt}
\noindent {\bf Statistical isotropy:}  Once statistical homogeneity has been secured, in the form of Eq. (\ref{shcond}), we ask about the invariance under spatial rotations (i.e. isotropy).  Of course again, the perturbation map is not isotropic, but it may be that the probability distribution function governing $\beta({\bf x})$ is, which is called statistical isotropy. This means that the $n$-point correlators in real space are invariant under rotations in space, i.e.
\begin{equation}
\langle \beta(\tilde{\bfx}_1) \beta(\tilde{\bfx}_2) ... \beta(\tilde{\bfx}_n) \rangle = \langle \beta({\bfx_1}) \beta({\bfx_2}) ... \beta({\bfx_n}) \rangle \,,
\end{equation}
where ${\bf \tilde{x}}_i = \mathcal{R} \ {\bf x}_i$, $\mathcal{R}$ being a rotation operator. To satisfy the above requirement, the $(n-1)$-spectrum must satisfy the condition
\begin{equation}
M_\beta (\tilde{\bfk}_1, \tilde{\bfk}_2, ... , \tilde{\bfk}_n) = M_\beta ({\bfk_1}, {\bfk_2}, ... , {\bfk_n}) \,, \label{sicond}
\end{equation}
where the tildes over the momenta represent as well a spatial rotation, parameterized by $\mathcal{R}$, in momentum space. This condition has more explicit consequences in the spectrum (1-spectrum) and the bispectrum (2-spectrum):
\begin{eqnarray}
M_\beta ({\bfk_1}, {\bfk_2}) &\equiv & P_\beta ({\bfk_1}, {\bfk_2}) = P_\beta (k) \,, \label{sicond1} \\
M_\beta ({\bfk_1}, {\bfk_2}, {\bfk_3}) &\equiv & B_\beta ({\bfk_1}, {\bfk_2}, {\bfk_3}) = B_\beta (k_1,k_2,k_3) \,, \label{sicond2}
\end{eqnarray}
where in the first line $k = |{\bfk_1}| = |{\bfk_2}|$, and in the second line $k_i = |{\bfk}_i|$. Starting from the trispectrum (3-spectrum), the condition in Eq. (\ref{sicond}) about statistical isotropy in all the higher-order $(n-1)$-spectra cannot be reduced to similar conditions to the ones in Eqs. (\ref{sicond1}) and (\ref{sicond2}), so that the minimal way of parameterizing the $(n-1)$-spectra (with $n \geq 4$) will always be in terms of all the $n$ wavevectors. The scalar nature of $\beta({\bf x})$ is very important since, if it were a vector or a tensor, there would not be a way to make the $n$-point correlators in real space invariant under spatial rotations. In those cases, we relax the meaning of statistical isotropy and establish that it is present if the $(n-1)$-spectra of the scalar pertubations that multiply the respective polarization vectors or tensors satisfy Eq. (\ref{sicond}).

\section{The separate universe assumption and the $\delta N$ formalism}

\vspace{3pt}
\noindent {\bf The separate universe assumption:} This assumption refers to the behaviour of the Universe after smoothing on a specified comoving scale $k^{-1}$, during the super horizon era $k \ll aH$.  It states that the spatial gradients, at most of order $k/a$, are negligible, which actually means that the Universe at each comoving location behaves as if it were homogeneous (Lyth \& Liddle, 2009).  Each smoothed region about each comoving location is then regarded as a separate homogeneous universe.  By virtue of this, the form of the equations for the dynamical quantities in each separate universe is the same as for the unperturbed quantities.  The separate universe assumption is a powerful tool for dealing with perturbations in the very early Universe and has become one of the most employed methodologies as alternative to the standard cosmological perturbation theory.

\vspace{3pt}
\noindent {\bf The $\delta N$ formalism:} The $\dn$ formalism (Dimopoulos et. al., 2009; Lyth \& Liddle, 2009) provides a powerful method to evaluate the primordial curvature perturbation $\zeta ({\bf x}, t)$ in terms of the perturbations of the fields a few Hubble times after horizon crossing $t_{*}$ (corresponding to a flat slicing), and the derivatives of the unperturbed number of $e$-foldings $N(t, t_{*})=\int_{t_{*}}^t  H(t')dt'$ with respect to the unperturbed fields evaluated at $t_{*}$.
According to this formalism, once the separate universe approach has been invoked, and a comoving threading has been established, the value of $\zeta$ in a uniform energy density hypersurface at the final time $t$ is given by the perturbation in the time integral of the local volume expansion rate $\theta$ along a curve starting at an initial flat hypersurface at the time $t_{i}$:
\begin{equation}
\zeta(\bfx,t) \equiv \delta N (\bfx,t, t_i) - \langle \delta N (\bfx,t, t_i) \rangle \,.
\end{equation}
Here, the bracket notation means a ensemble average (which corresponds to a spatial average if there is statistical homogeneity). In many inflationary scenarios, the number $N$ of $e$-foldings depends only on the values of the fields at $t_*$ so we can write the curvature perturbation as an expansion in the perturbations of the fields at this time.  Assuming the presence of just one scalar field and one vector field (the generalization to more fields is straightforward), we have:
\begin{eqnarray}
\zeta(\bfx , t)\equiv\delta N (\phi(\bfx),A_i(\bfx),t) &=& N_\phi \delta\phi + N_i\delta A_i+\frac{1}{2}N_{\phi\phi}
(\delta\phi)^2+
N_{\phi i}\delta\phi\delta A_i + \nonumber \\
&& +\frac{1}{2}N_{ij}\delta A_i \delta A_j + \ldots \,,\label{dScsvec}
\end{eqnarray}
where
\begin{equation}
N_\phi\equiv\frac{\partial N}{\partial \phi}\,,\quad
N_{\phi\phi} \equiv\frac{\partial^2 N}{\partial \phi^2}\,,\quad
\quad N_{\phi i}\equiv\frac{\partial^2 N}{\partial A_i\partial\phi}\,, \quad \rm{etc}. \,,
\end{equation}
are the derivatives with respect to the scalar $\phi$ and the spatial components of the vector field ${\bf A}$. 

\section{FRW background expansion implies statistical isotropy} \label{s1}

We will make the following assumptions:

\begin{enumerate}
\item The action is such that the FRW metric is an attractor in the background. \label{a1}
\item The fields involved are just scalar and/or vector fields. \label{a2}
\item The background expansion is FRW type during the whole time spanning from the beginning of inflation to the time when the curvature perturbation $\zeta$ is evaluated. \label{a3}
\end{enumerate}

By invoking the separate universe assumption, the form of the equations for the dynamical quantities at each comoving location is the same as for the unperturbed quantities.  Thus, the field perturbation equations in momentum space do not depend explicitly on ${\bf k}$.  They depend on time, thus on $k_*$, but do not depend on the direction of ${\bf k}$.  Their solutions are, therefore, independent of the direction of ${\bf k}$ except for the set of initial conditions $\{\alpha^n_{\bf k}\}$.  However, the field perturbations are evaluated in the flat slicing, so, taking into account the assumption \ref{a3}, the whole perturbed metric in this slicing is actually FRW which is conformally equivalent to Minkowski.  This has as a consequence that the set of initial conditions, when quantizing, can be written as $\{\hat{\alpha}^n_{\bf k} = \alpha^n_k \hat{a}^n_{\bf k}\}$ where $\hat{a}^n_{\bf k}$ is the respective annihilation operator.  Following the usual procedure to calculate the power spectrum of the field perturbations, this implies that the latter are statistically isotropic (and statistically homogeneous) at the level of the power spectrum (since the $\alpha^n_k$ do not depend on the direction of ${\bf k}$).  This is valid for all the relevant cosmological scales since the time of horizon exit for each of them is contained in the period of time defined in the assumption \ref{a3}.

Now, if the background expansion is FRW type, the fluid that fills the universe is perfect;  therefore, by invoking assumption \ref{a2}, the field configuration must be given by any number of scalar fields and/or multiple randomly oriented copies (just one is enough) of a triad of mutually orthogonal vector fields with the same norm.  This must be accomplished during the whole period of time defined in assumption \ref{a3} \footnote{This has as a consequence that the action must be consistent with the isotropy during the referred period of time, e.g., the masses for the members of each triad must be equal (although not necessarily the same masses among triads).}.  Thus, by employing the $\delta N$ formalism, we obtain (Gómez \& Rodríguez, 2013)
\begin{equation}
P_{\zeta}(\textbf{k}_{1})=P_{\zeta}^{iso}(k_{1})[1+g_\zeta^{1}(\hat{\textbf{k}}_{1}\cdot
\hat{\textbf{N}}_1)^{2}+g_\zeta^{2}(\hat{\textbf{k}}_{1}\cdot
\hat{\textbf{N}}_2)^{2}+g_\zeta^{3}(\hat{\textbf{k}}_{1}\cdot
\hat{\textbf{N}}_3)^{2}] \,,\label{b14}
\end{equation}
where 
\begin{equation}
P_{\zeta}^{iso}(k_1)=(N_{\phi})^{2}P_{\delta\phi}(k_{1})+(N_{i}^{1})^{2}P_{+}^{1}(k_{1})+(N_{i}^{2})^{2}P_{+}^{2}(k_{1})+(N_{i}^{3})^{2}P_{+}^{3}(k_{1}) \,,\label{b13}
\end{equation}
and
\begin{equation}
g_\zeta^{n}=\frac{(N_{i}^{n})^{2}[P_{long}^n(k_1) - P_{+}^{n}(k_{1})]}{P_\zeta^{iso}(k_1)} \,,\label{b12}
\end{equation}
$P_{+}^n(k)$ being the parity even spectrum of the $n$-th vector field, $P_{long}^n(k)$ being the longitudinal spectrum of the $n$-th vector field, and
\begin{equation}
\hat{\bf{N}}_n = \frac{{\bf N}_n}{|{\bf N}_n|} \,,
\end{equation}
where ${\bf N}_n$ is the vector formed by the derivatives of $N$ with respect to the each spatial component of the $n$-th vector field.  However, because of the symmetries of the field configuration, we have
\begin{eqnarray}
\hat{{\bf N}}_1 &=& \pm \hat{\bf i} \,, \\
\hat{{\bf N}}_2 &=& \pm \hat{\bf j} \,, \\
\hat{{\bf N}}_3 &=& \pm \hat{\bf z} \,, \\
(N_i^1)^2 &=& (N_i^2)^2 = (N_i^3)^2 \,, \\
 P_{+}^1 &=&  P_{+}^2 =  P_{+}^3 \,, \\
  P_{long}^1 &=&  P_{long}^2 =  P_{long}^3 \,, 
\end{eqnarray}
and, therefore,
\begin{equation}
g_\zeta^1 = g_\zeta^2 = g_\zeta^3 = g_\zeta\,.
\end{equation}
Thus,
\begin{eqnarray}
P_{\zeta}(\textbf{k}_{1}) &=& P_{\zeta}^{iso}(k_{1})[1+g_\zeta [(\hat{\textbf{k}}_{1}\cdot
\hat{\bf i})^{2}+(\hat{\textbf{k}}_{1}\cdot
\hat{\bf j})^{2}+(\hat{\textbf{k}}_{1}\cdot
\hat{\bf z})^{2}]] \,, \\
&=& P_{\zeta}^{iso}(k_{1}) [1+g_\zeta] \,,
\end{eqnarray}
where the last step is valid because of the director cosines property of a unit vector.  We conclude then that the power spectrum of $\zeta$ is actually independent of the direction of ${\bf k}_1$, rendering the two-point correlator of $\zeta$ in real space statistically isotropic.

\section{Non-FRW background expansion implies statistical anisotropy} \label{s2}

If the background metric is not FRW at some time, then the background metric is not FRW at $t_*$ because of assumption \ref{a1}.  Thus (Gómez \& Rodríguez, 2013),
\begin{eqnarray}
P_{\zeta}(\textbf{k}_{1}) &=& P_{\zeta}^{iso}(k_{1})[1+ \tilde{g}_{\delta \phi} (\hat{\textbf{k}}_{1}\cdot
\hat{\textbf{d}}_{\delta \phi})^{2} + \tilde{g}_+^1 (\hat{\textbf{k}}_{1}\cdot
\hat{\textbf{d}}_+^1)^{2} + \tilde{g}_+^2 (\hat{\textbf{k}}_{1}\cdot
\hat{\textbf{d}}_+^2)^{2} + \nonumber \\
&& + \tilde{g}_+^3 (\hat{\textbf{k}}_{1}\cdot
\hat{\textbf{d}}_+^3)^{2} + \tilde{g}_N^{1}(\hat{\textbf{k}}_{1}\cdot
\hat{\textbf{N}}_1)^{2}+\tilde{g}_N^{2}(\hat{\textbf{k}}_{1}\cdot
\hat{\textbf{N}}_2)^{2}+\tilde{g}_N^{3}(\hat{\textbf{k}}_{1}\cdot
\hat{\textbf{N}}_3)^{2}] \,, \nonumber \\
\end{eqnarray}
where none of the $\tilde{g}$, except perhaps for the $\tilde{g}_N$ (if the background metric is again FRW at the time when $\zeta$ is evaluated),  are the same, and, in contrast, all the $\hat{\bf d}$ directions are the same and equal to the preferred direction at $t_*$. We conclude then that the power spectrum of $\zeta$ does depend in this case on the direction of ${\bf k}_1$, rendering the two-point correlator of $\zeta$ in real space statistically anisotropic.

\section{Conclusions}
When we take sections  \ref{s1} and \ref{s2} together, they show that, under the established assumptions, the FRW background expansion is equivalent to the statistical isotropy of $\zeta$ in connection with its two-point correlator in real space during some period of time that contains primordial inflation.

\acknowledgments This work was supported by COLCIENCIAS grant number 110656933958 RC 0384-2013 and by COLCIENCIAS - ECOS-NORD grant number RC 0899-2012.  Y.R. acknowledges VIE (UIS) for the financial support through its mobility programme.


\begin{references}

\reference Ade, P. A. R. et. al. 2014, Astron. Astrophys. 571, A23.

\reference Bennett, C. L. et. al. 2013, Astrophys. J. Suppl. Ser. 208, 20.

\reference Dimopoulos, K. et. al. 2009, JCAP 0905, 013.

\reference Gómez, L. G. \& Rodríguez, Y. 2013, AIP Conf. Proc. 1548, 270.

\reference Lyth, D. H. \& Liddle, A. R. 2009, ``The primordial density perturbation'', Cambridge University Press, Cambridge (UK).




\end{references}
\end{document}